# Search for deviations from the ideal Maxwell-Boltzmann distribution for a gas at an interface

P. Todorov[1], J.C. de Aquino Carvalho[2,3], I. Maurin[2,3], A. Laliotis[2,3], D. Bloch[3,2]*

[1] Institute of Electronics–Bulgarian Academy of Sciences, 72 Tsarigradsko chaussee blvd., Sofia, Bulgaria
[2] Laboratoire de Physique des Lasers, Université Paris13-USPC, 93430 Villetaneuse, France
[3] CNRS, UMR7538, Laboratoire de Physique des Lasers, 99 av JB Clément 93430 Villetaneuse, France

**ABSTRACT**

The isotropic Maxwell-Boltzmann (M-B) velocity distribution is the accepted standard for a gas at thermal equilibrium, with the Doppler width considered to deliver a very precise measurement of the temperature. Nevertheless, the physical nature of the surface, and the atom-surface interaction, in its long-range (van der Waals type) regime as well as in its short-range regime leading to adsorption/desorption mechanisms, are far from the ideal situation describing the foundation of gas kinetics. Through the development of vapor spectroscopy at an interface, a high sensitivity to atoms flying nearly parallel to the gas interface is obtained, and deviations to M-B distribution could have observable effects, even affecting the ultimate limits to resolution. We report here on the development of an experiment involving a dedicated set-up of spatially-separated pump-probe experiment in a narrow cell. A first series of investigation could not evidence a deviation for Cs atom velocities at a nearly grazing incidence (typically 1.5 -5°). We discuss various technical improvements to increase the sensitivity to atoms flying parallel to the surface, along with specific spectroscopic information that may be collected. Alternately, the comparison of standard selective reflection spectroscopy with simultaneous volume spectroscopy may be a source of complementary information. At last, we discuss how the Boltzmann energy distribution, among molecular or atomic levels, may become sensitive to specific thermal exchanges at the surface, in an equilibrium situation.

## 1. INTRODUCTION

The gas kinetics theory was extremely successful as early as the end of the XIX[th] century, and contributed, along with the development of chemistry, to impose the idea of the molecular structure of matter. In these times, the size of this "atomic" structure was a difficult problem. Not surprisingly, the description of the boundary enclosing the gas was very elementary, and was limited to ideal interfaces.

### 1.1  Specular collisions, scattering, and accommodation on surfaces

An excellent introduction to the history of the gas kinetics, with its initial assumptions which cannot hold in view of what is known about surface desorption properties, is given in the review article by Comsa and David [1]. It recalls that the Maxwell-Boltzmann velocity distribution, aside from a kinetic energy distributed according to Boltzmann law of energy equipartition, assumes isotropy of the velocity distribution, as a result of numerous random atom collisions.

An interesting point is that the original Maxwell description considers the possibility of two very different behaviors for atom impinging the surface: a scattering behavior, whose probability is traditionally named *f*, while (1-*f*) is the probability for a specular collision. If what is currently known about atom-surface interaction at a quantum level makes us reluctant to consider specular reflection as a common behavior, the idea was naturally inherited from classical mechanics and billiard models. Allowing a scattering behavior for a fraction *f* of the colliding molecules is an interesting intuition, but this is oversimplified with respect to all the detailed analysis of microscopic mechanisms occurring at a real surface. Note that in the presence of thermal disequilibrium, as it occurs for a surface whose temperature differs from the one of the surrounding gas, an accommodation coefficient is often used, notably in engineering sciences such as aeronautics: the mean kinetic energy, for particles having bounced from the surface, results from an accommodation-weighted average, between the temperature of the gas, and the surface temperature.

*daniel.bloch@univ-paris13.fr,  phone +33 149403390

To analyze the problems occurring close to the boundary, the concept of a specific layer was introduced (Knudsen) to describe the region where the molecular "mean free path", derived from the gas kinetics, becomes irrelevant owing to the "collisions" with the surface.

### 1.2 The cos θ law and desorption processes

As a consequence of the isotropy of the M-B velocity distribution, the flux of the gas particles leaving the surface has to follow a "cos θ" law, with θ the angle between the particle trajectory and the surface normal. It was notably recognized [1] by Clausing that this implies that the "cos θ" must apply as well for desorbing particles —or for evaporating particles, when the main mechanisms retained for the scattering was a condensation/evaporation process. Clausing noted that the origin of such a "cos θ" law was not justified, but it was found to agree with experiments of the early times[2].

### 1.3 Real surfaces and the limits of the M-B description

With the development of molecular beam techniques, it appeared later on that the geometry for scattering and desorption processes was much richer[2,3]. Peaked angular velocity distributions were observed, along with more complex behaviors, such as atomic diffraction from solid surfaces[3]. This commonly leads to an emerging flux expanded in powers of cos θ, as $\Sigma_n\, a_n (\cos\theta)^n$.

In addition, the interaction energy between the particle and the surface has been evaluated. It reaches values comparable to, or largely exceeding, a fraction of eV, typically characterizing the "frontier" between physisorption and chemisorption. These values are well above the ~ 25 meV average kinetic energy of a gas particle at room temperature. Hence, molecules undergo considerable accelerations at short distances from the surface.

These effects, occurring at the microscopic scale, make the hypotheses involved in the M-B velocity distribution highly questionable. Nevertheless, the equilibrium situation assumes an integration over the whole boundary regions enclosing the free gas: this is much more delicate to analyze, with effective studies limited to very specific cases. The possibility to prepare an anisotropic velocity distribution by a specific tailoring of the surface seems to remain an open question.

### 1.4 Doppler-broadened and sub-Doppler spectroscopy as informative on the velocity distribution

A considerable success of laser spectroscopy and of the M-B distribution as well, is in the development of various techniques of Doppler-broadened and sub-Doppler methods, that all look to agree perfectly with the M-B predictions. In its pure "gas" version, neglecting any influence of the boundary on the macroscopic gas, linear Doppler-broadened absorption provides[4] a competitive measurement of the Boltzmann constant. Under this basic assumption of M-B law for the velocity distribution, it is now proposed to build metrological thermometers consisting of a gas at equilibrium, with the measurement provided by a spectrometer through the analysis of the Doppler width.

Numerous sub-Doppler spectroscopy techniques are also able to address selectively a given velocity group. The effective velocity distribution, inside a molecular beam, can be revealed by high resolution laser spectroscopy. Conversely, velocity-selective nonlinear spectroscopy, notably saturated absorption and related multi-level techniques, allows to select a narrow velocity group inside the gas of a cell. It is not our purpose here to recall the factors finally limiting the very high resolution of these techniques: let us just recall [5] that molecular velocities are currently selected with an accuracy in excess of $10^{-3}$ of the thermal velocity, as long as the homogeneous broadening is a small enough fraction of the Doppler-broadened width. The major requirements are a sufficient laser frequency stability, moderate saturation effects, and collimation of the beams (*e.g.* better than 1 mrad) to avoid any residual Doppler-broadening.

Spectroscopy of resonant gases at an interface has been largely developed, notably in our lab (see *e.g.* a review [6]) with a particular emphasis on techniques allowing a sub-Doppler resolution through a selection of atoms whose velocity direction corresponds to a grazing incidence. These techniques, among which selective reflection spectroscopy [6-14], and spectroscopy in a thin cell [15,16] –or extremely thin cell[17] –, have the ability to provide a sub-Doppler resolution with a single beam irradiation, as long as the irradiation is under normal incidence. Indeed, for atoms traveling nearly parallel to the surface, the Doppler shift is null or strongly reduced while interaction with the resonant light reaches an efficient (steady-state) regime: this is notably true even very close to the surface, either for atoms having just left the surface (assumed to be in the ground state), or for arriving atoms experiencing suddenly a large surface interaction. In practice, these techniques were mostly implemented on transitions of alkali metal vapor, for which the natural width commonly exceeds $10^{-2}$ of the Doppler width. Even in situations where the natural width could be slightly less than $10^{-2}$ of the Doppler width [7,8], the observed sub-Doppler width has always exceeded $2.10^{-2}$ of the Doppler width. It is only in recent experiments, on molecules at an interface, for experiments in the mid infrared [~ 10.5μm] that a factor slightly better than $10^{-2}$ has been recently obtained [9]. In this last case, the selection of slow molecules occurs at much larger distances than for a visible -near IR range spectroscopy, owing to the longer wavelength.

The use of the sub-Doppler features of thin cell spectroscopy was proposed[18] for frequency metrology purposes, extrapolating the demonstrated experimental results to narrow atomic transitions. It is then assumed that the distribution of (normal component) velocity is truly M-B, *i.e.* flat around $v_\perp = 0$, even at the level of $\sim 10^{-4}$ of the thermal velocity. For the intercombination line of Ca, this implies[18] a selectivity well below 10 cm/s, comparable to the velocities reached in laser cooling techniques —but in this last case, far away from any surface. As noted earlier[15], selecting very slow atoms actually addresses fundamental questions on the gas kinetics at a boundary, defined in the lab frame.

Finally, let us recall that there is only a single experimental report [19] having probed the cos θ law through laser spectroscopy in a thermal vapor. With the limited laser frequency resolution of these early times, it demonstrated only that the cos θ law is in an acceptable agreement, while a trial law cos² θ is invalid. It is in this spirit that we have pursued a research line on this topic, with dedicated experiments [20] performed in the IE-BAS laboratory.

## 2. EXPERIMENTS WITH A DEDICATED SET-UP

### 2.1 Principle

The principle of the experiments[20], already described in an earlier conference of this series [21], follows a brief attempt [22] in the line of the initial spectroscopy experiments in a thin cell [15]. Essentially, one performs a spatially-separated pump-probe saturated absorption experiment, with both beams under normal incidence with respect to the parallel windows of the thin cell. The test consists in analyzing if the number of "slow" atoms (relatively to the normal to the surface, *i.e.* atoms flying nearly parallel to the wall) survives according to the cos θ law, when one increases the pump-probe separation. In the experimental set-up, the pump is ring-shaped, thanks to a dedicated optical system (circular grating + axicon), and the probe is at the center of the ring-shaped pump (fig. 1). For alkali atoms (Cs in our case), the hyperfine optical pumping in the ground state is long-lived, so that, if the cos θ law is verified, the density of pump-marked atoms reaching the probe region should simply decrease as the inverse of the pump-probe distance.

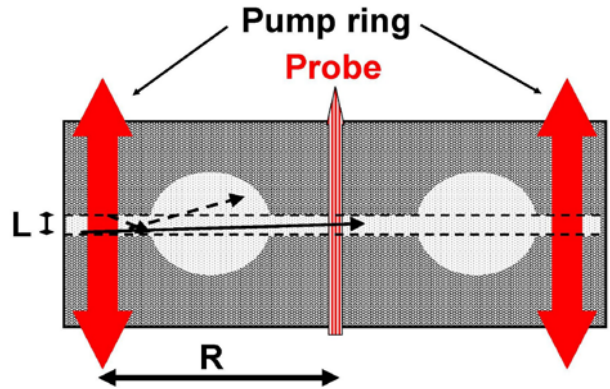

Figure 1: Scheme of the experiment (top-view). The cylindrical thin cell is irradiated under normal incidence with a ring-shaped pump (diameter R), and the probe goes along the axis of the pump ring. In the set-up we used, the probe beam counterpropagates the pump beam, only for a better elimination of stray signals. Only pumped atoms with a nearly parallel velocity (full line arrow) can reach the probe region. The "thin" cell, of an internal thickness L, may be partially etched between the pump and the probe, to avoid the counting of atoms (dashed line arrow) bouncing onto the surface.

Two major causes may affect this prediction derived from the M-B distribution assumption : on the one hand, too "slow" atoms may not survive, as due to surface attraction or intrinsic roughness of the microscopic surface, so that the decrease with the distance is enhanced the decrease would be smaller if, unexpectedly, slow atoms are more numerous than expected[14]). On the other hand, pumping memory may generate signals featuring "false positive", with the probe detecting "slow" atoms, induced by a collision in the separation zone between the pump and the probe.

## 2.2 Experimental results

Our complete report[20] on the experiments shows signals observed for a pump-probe separation up to R = 2 mm, in a L = 60 µm thin cell. The pump power is sufficient to induce full saturation and a signature of the difference between a situation with a genuine separation zone, and a geometry with a partial pump-probe overlap, is the absence of cross-over resonances on the saturated absorption spectrum, as long as there is no residual overlap —note that despite the relatively small thickness of the cell, crossover resonances[18] remain visible under a partial pump-probe overlap.

The key result, once the method is validated (pump intensity strong enough to saturate the pumping process, probe linehapes independent of the pump-probe separation, ...) is that the for each hyperfine component, the absorption on the probe, limited to its pump-induced component (whose effect is actually an absorption *reduction* through saturation) decreases with the inverse of the pump-probe separation –see fig. 2. From this demonstration experiment, one can conclude that for the selected (normal) velocities (given by R/L.$u_{th}$, with $u_{th}$ ~ 200 m/s) down to ~ 5-20 m/s, the M-B distribution remains acceptable. This is not too surprising relatively to current observations a variety of experiments involving velocity-selective spectroscopy –although they were not performed in view of a systematic analysis. Nevertheless, gaining an order of magnitude in the selection of slow atoms appears feasible[20]. This would allow to reach a level where deviations to the M-B law are expected because of the surface interaction, and residual surface roughness. It is expected that the observation of such a deviation should strongly depend on the nature of the surface.

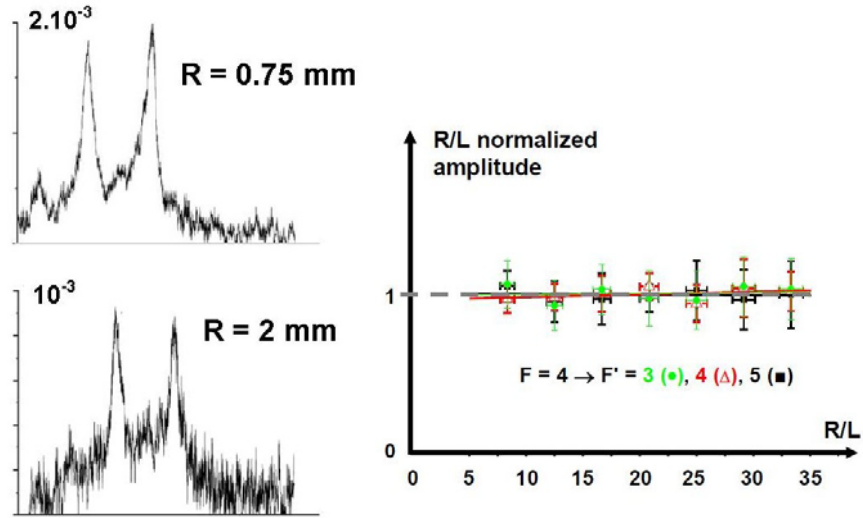

Figure 2: The left part shows two frequency spectra on the Cs $D_2$ line [F = 4 → F' = {3,4,5}] for the two respective separation R = 0.75 mm ad R = 2 mm. Note the decrease of the amplitude for the larger separation. The residual broad background is subtracted in the analysis of the peak amplitude. The right part shows, for the various hyperfine components, the $R^{-1}$ normalized amplitude of each hyperfine component when investigating the R/L dependence —see the original report[20] for more details.

## 2.3 Towards a second generation of experiments

More up-to-date equipments would remove several limitations of the experiments reported above. In particular, up to now, the experiments were performed with a single tunable laser, in a free-running mode. Its linewidth, although much smaller than the Doppler broadening, largely exceeds the Cs resonance natural width. A detailed analysis of the probe absorption lineshape could be performed with two independent lasers. This would allow a more precise evaluation of the residual effect of atom-atom collisions. Relatively to fig. 1, the pump-induced probe absorption should be narrower, and any broadening would sign the contribution of collisions, and of unwanted atoms, which are not in free flight form the pump region. Moreover, analyzing the probe absorption lineshape may allow to retrieve the velocity distribution through a deconvolution of the Lorentzian absorption. Figure 3 shows that significant signatures in the lineshape are predicted for rather low R/L value. Moreover the pumping of slow atoms could even be detected on narrower transitions., *e.g.* the 455 nm line of Cs, possibly resolving velocitysmaller than $u_{th}$/100. Such a method, based upon a lineshape analysis, would be more effective than the one of our initial demonstration, based upon the R/L evolution of the *peak* amplitude.

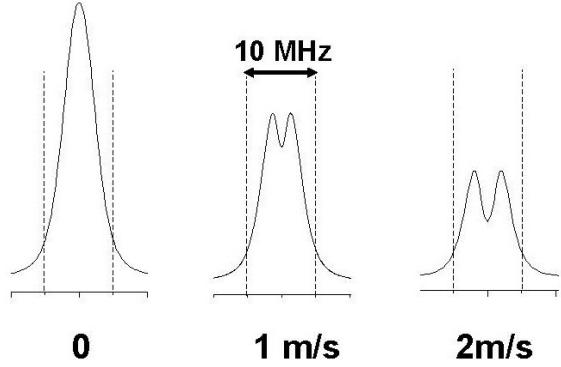

Figure 3: Calculated lineshapes for the (pump-induced) probe transmission for R/L = 40, a Doppler width $ku_{th}$ = 200 MHz, and a homogeneous transition width 5 MHz. A genuine M-B distribution is assumed (left), or a truncated one, with minimal normal velocity $|v^\perp|$ 1m/s (center), or 2m/s (right). These lineshapes result from a convolution between the absorption –sensitive to the Doppler-shift–, and the velocity distribution of the pumped atoms reaching the probe region —see fig. 2 of the original report[20]. When the slowest atoms are missing, the drop in the signal amplitude goes along with a dip, associated to a residual (blue and red) Doppler shift (from P. Todorov, unpublished).

Aside from an improved diagnostics of possible stray effect in the signature of atoms freely moving from the pump to the probe, another reason for getting two independent frequency-narrowed lasers (*e.g.* extended cavity lasers) is that only lasers with a good frequency-stability allow accumulation data. Up to now, results were obtained only for rather short scans (few minutes or less): frequency drifts make longer scans irreproducible, and the data were only stored in the memory of an oscilloscope, instead of a dedicated recording system. The key for a more refined velocity selection is obviously a longer integration time, to count sensitively the rarefied distribution of atoms at very near grazing incidence.

At last, atom-atom collisions can be a major limitation to the ability of detecting long flight of atoms at grazing incidence. A pumped atom bouncing several times without hyperfine relaxation is counted as a "slow" atom if it finally reaches the probe region with a grazing velocity. Temporal selectivity in the pump modulation/probe detection process could discriminate a large fraction of these unwanted atoms because through scattering and random velocity redistribution, the longitudinal walk towards the probe beam is very inefficient owing to the specific cell geometry. In addition, a specific design[20] of the cell (fig.1), with an etching of the windows between the pump-probe separation region, would prohibit most of the pumped atoms to reach the probe region if they have undergone a collision.

## 3. COMPARING SPECTROSCOPY IN THE VOLUME AND AT AN INTERFACE

In the course of experiments[8] aiming to measure the temperature dependence[7,8] of the fundamental long-range atom-surface interaction [near-field van der Waals (vW) regime of the Casimir-Polder interaction], we investigated in detail Selective Reflection (SR) spectroscopy at a sapphire interface on the Cs $D_1$ resonance line ($\lambda$=894 nm, natural width 4.6 MHz), in a low density cell. Here we report on an intriguing result[8] in the comparison between volume spectroscopy with saturated absorption (SA) and frequency-modulated (FM) selective reflection (SR) under normal incidence. Both techniques are indeed velocity-selective Doppler-free techniques, selecting the same narrow contribution, with FM SR specifically selecting the contribution of atoms flying nearly parallel to the wall.

### 3.1 Linear SR spectroscopy at the interface on the Cs $D_1$ line

The experiment was performed with a commercial extended cavity-laser, frequency tunable, whose linewidth is ≤ 2 MHz, and the p-p excursion of the applied FM is ~2 MHz. Our choice was to look for the minimal width of the SR spectroscopy on the resonance $D_1$ line. SR spectroscopy had been investigated on this Cs resonance line[10] only in a defective cell, filled with impurities: the large collision broadening was shown to have only a small influence for the measurement of the atom-surface interaction. Here, with a Cs cell nicely operating, we have tried to reach the minimal width for SR spectroscopy: this implies a low Cs density, resulting into a weak amplitude signal, but this is compensated for by a longer integration time (data accumulates over successive frequency scans, with a re-centering over the resonance, in order to eliminate the long term frequency drift). In addition, the irradiation intensity is kept low, to minimize the saturation effects, occurring easily with narrow widths.

As usual[11-12], the FM SR lineshapes exhibit a vW-induced shift and distortion relatively to Doppler-free spectroscopy in the volume (SA in an auxiliary cell). The observed lineshape is truly independent of the Cs density because the low operating temperature of the Cs reservoir is as low as $\leq 60$ °C (*i.e.* ~$10^{12}$ at.cm$^3$) —collision broadening starts to be observed for a reservoir temperature $\geq 90$-$100$°C. The minimal width for this lineshape was carefully analyzed[8] with a robust model [12] taking into account the vW interaction. It allows excellent fitting of the lineshapes (see fig. 4), with the fitting strength of the vW interaction compatible with the predicted values. Nevertheless, a remarkable point is that the linewidth extracted from the model is ~8.5 MHz. Note that this extracted minimal linewidth –with respect to the low density– is not exactly equal, although close, to the distance between the two peaks of the nearly dispersive lineshape. This experimental minimal value, slightly exceeding the expected value, is by the way similar to the one ($\geq 8$ MHz) obtained in many occurrences[11] on the Cs $D_2$ line (natural width: 5.3 MHz), at the limit of a very low collision broadening —and although systematic investigations were not conducted on the $D_2$ line.

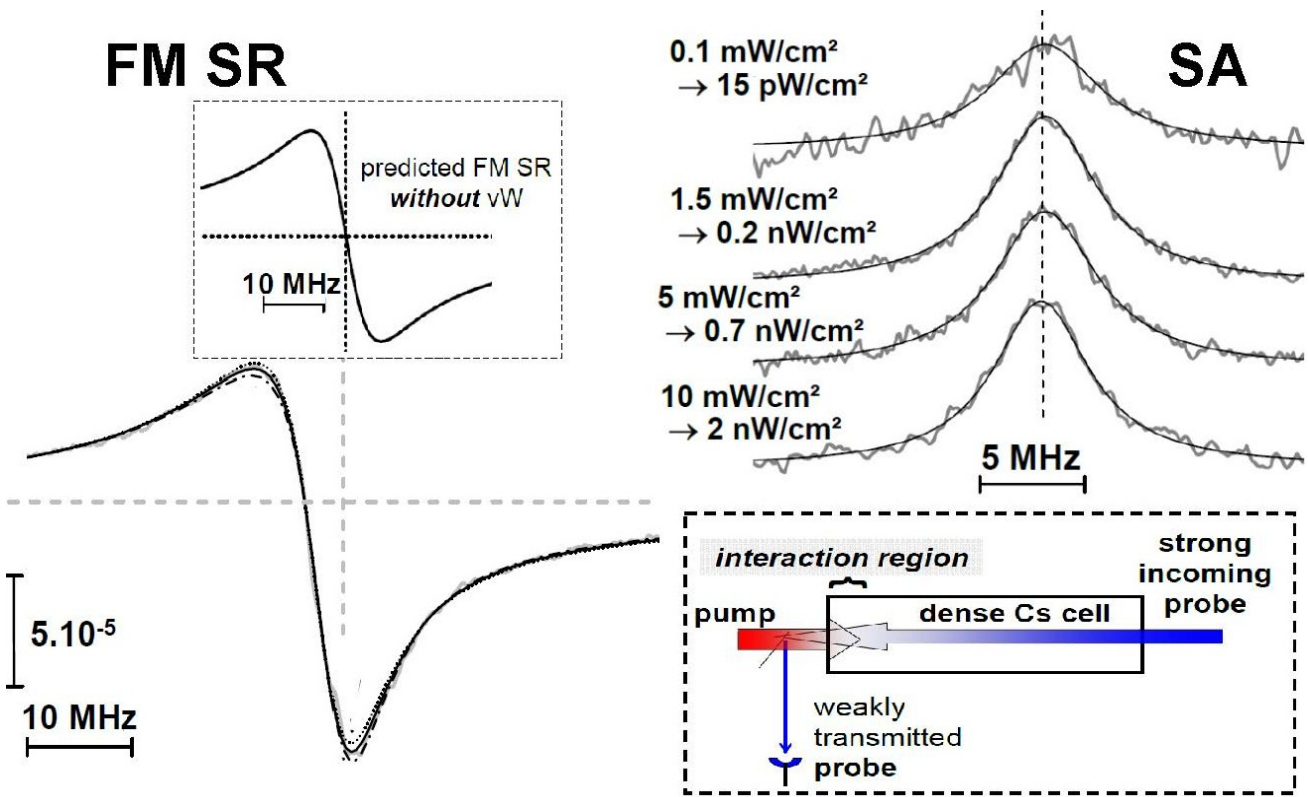

Figure 4: Comparison between: (left) FM Selective Reflection (SR) spectroscopy and: (right) Saturated Absorption (SA) spectroscopy on the Cs $D_1$ line (F = 4 → F' = 3 hyperfine component) in the same very low-density vapor cell (Cs reservoir at ~ 60°C, *i.e.* 30 µTorr, and cell body temperature ~ 150°C). The (FM) SR lineshape deviates from a perfect antisymmetric lineshape because of the vW attraction, inducing shift and asymmetry. Fits with the vW attraction included nearly coincide with the experimental spectra —an optimal and approximate fittings are represented. For all these fits, the extracted homogeneous width is 8.5MHz, with the reconstructed lineshape (in the absence of the vW interaction) shown in the insert (top). The SA spectra are recorded with an input pump intensity 1µW/cm² (at the SR window), and various probe intensities (as shown). Because the counterpropagating probe propagates over the 8 cm length cell, it is strongly attenuated —see the principle scheme in the insert below. The SA spectra are fitted with a 6 MHz linewidth Lorentzian, independently of the input probe intensity. The probe power in the effective region of interaction is very weak, slightly exceeding the weak measured transmitted intensity –value indicated following the arrow. More details on the spectra can be found in figs. 6.5 and 6.16 of the report[8].

### 3.2 Saturated absorption spectroscopy in the same vapor used for FM SR

A direct comparison between SR spectroscopy (at a gas interface), and Doppler spectroscopy in the volume of the same gas, had not been reported to our knowledge. This is because SR spectroscopy requires the light vapor interaction to be strong enough to be produce a signal over a typical distance of one (reduced) wavelength ($\lambda/2\pi$); coversely, Doppler-free spectroscopy methods are well modeled for the "optically thin" approximation, requiring a "macroscopic cell" (typical thickness >> 100 $\lambda$), *i.e.* far away from the opacity typically needed to perform SR spectroscopy.

For the very same cell used in the previous section for spectroscopy at an interface, we have implemented a dedicated set-up[8], yielding specifically the SA spectrum close to the interface used for the SR experiment. The trick is in using a relatively weak (amplitude-modulated) pump beam, entering at the interface of interest, which attenuates on a short distance (the attenuation length is a few mm for a Cs vapor temperature ~ 50°C). Hence, the pump-induced saturation effects occur only close to the interface. Over its total propagation (8 cm for our cell), the probe beam undergoes a very strong attenuation ($\geq 10^{-4}$) despite a strong saturation intensity where it is launched, at the opposite window. The intensity of the transmitted probe is easily detected with a low-noise detector, and the key point is that, where the pump beam is efficient, the probe beam itself has reached a non-saturating intensity.

The major result is that the pump-induced absorption exhibits a lineshape whose linewidth is ~ 6 MHz. Despite experimental conditions seemingly unusual only at first sight, the experiment is nothing else than a saturated absorption performed with low intensity beams, and with negligible collisions: the 6 MHz linewidth is easily fitted with an absorptive Lorentzian. The slight excess for the measured linewidth, with respect to the 4.6 MHz natural width, can be probably results from the combined effect of the laser linewidth, the broadening by the FM applied to the laser, and possible residual impurities in the cell. Strikingly, this 6 MHz width is significantly smaller than the one (8.5 MHz) resulting from the SR lineshape analysis.

To confirm that this excess linewidth for the SR experiment is not a kind of artifact induced by the major difference between the two lineshapes (a distorted dispersion *vs.* a Lorentzian), we also analyzed[8] the saturated absorption spectrum in its FM version. It yields indeed an antisymmetric derivative of a Lorentzian, with a peak-to-peak width typically smaller (~ 4 MHz), by the expected factor $\sqrt{3}$, ruling out an effect of the FM, or of the laser jitter broadening for the SR spectrum.

### 3.3 Revisiting selective reflection specta with respect to the actual distribution of "slow" atoms

The intriguing point in the difference between the two observed widths is that SA is known to correspond to a velocity selection in the free space: when pump and probe frequencies are not at the same frequency, the velocity selection and the linewidth remain as narrow, despite a shift in the resonance frequency (simply analyzed in a moving frame). This is also why, in SA spectroscopy, the crossover resonances for a three-level system are associated to a specific (not null) longitudinal velocity. At the opposite, the theoretical calculation[13] conducting to the sub-Doppler linewidth of FM SR essentially relies on the assumption that the (normal) velocity distribution is flat around $v_\perp = 0$. This symmetry around $v_\perp = 0$ justifies that in linear SR spectroscopy, the contribution of arriving atoms (integrated over a half M-B distribution) is identical[12,13] —and not frequency-symmetric— to the one of atoms departing from the surface —also integrated over a half M-B distribution. Here, the observed difference between volume spectroscopy, and SR spectroscopy at the interface, suggests that this traditional hypothesis is worth to be seriously checked.

A related comparison, of interest if performed in the very same conditions, would be between the linear absorption (in the neighborhood of an interface), and SR (or FM SR) at the interface, which is also a linear technique of spectroscopy. Indeed, the amplitude of the FM SR signal is expected to be the identical (within some prefactor) to the one of the linear absorption over an absorption length $\lambda$. However, the FM SR amplitude is proportional to the number of slow (normal) atoms, while linear absorption depends on the total number of atoms in the vapor. Note as a subtle point that, although at a given frequency, the absorption is limited to the absorption of the resonant velocity group, this "velocity group" itself has a width defined by the homogeneous width of the transition: hence, assuming the large Doppler width approximation, the absorption remains independent of the homogeneous width. A velocity distribution violating the M-B prediction around $v_\perp = 0$ should hence lead to unexpected relative amplitudes, when comparing the linear absorption, and the FM SR signal.

In the same spirit, the predicted lineshape in direct SR spectroscopy (without FM) combines a mix of Doppler-broadened dispersion, and a "logarithmic" singularity[13] —in the limit of the "infinite" Doppler width. To our knowledge, this direct SR lineshape has never been investigated in detail, nor the corresponding experiments truly fitted with theory. Rather, a qualitative agreement, and the evidence of Doppler-free FM SR (*i.e.* the frequency-derivative of the sub-Doppler logarithmic singularity) have been demonstrated. The technical reason for this lack of detailed investigation is the low contrast (relatively to non resonant reflection) of the SR signal, especially for a low-density vapor. However, such a comparison has the ability to check intimately the predictions associated to a genuine elementary M-B distribution.

## 4. THERMAL SURFACE EMISSION AND ENERGY DISTRIBUTION FOR MOLECULES

Our major investigation here has been concerned by testing the isotropy assumption in the M-B distribution, close to an interface. Nevertheless, resonant couplings between the molecules of the vapor and surface modes, are susceptible to affect also the distribution of internal energy for the molecules of the vapor. We have demonstrated that the long range vW atom-surface potential can notably depend[7,8] on the surface temperature. This temperature-induced behavior should go along with a *real* selective transfer, between the internal energy of the atom and the surface mode. Such a Förster-like surface-induced energy transfer had been seen on highly excited Cs atom, as a surface-induced quenching[23] on a specific channel. It was observed in the absence of temperature effects (*i.e.* T = 0 limit); it is now searched for[8], in a reverse scheme, with the atom gaining excitation through the near-field thermal emission of the (hot) surface.

At thermal equilibrium, it is of interest to consider a molecular gas –rather than an atomic gas–, because molecular vibration modes typically fall in the infrared range, possibly exhibiting coincidences with polariton surface modes. Convenient materials such as sapphire and SiC exhibit surface modes[6-8], respectively at 12.2 µm, and ~10.5 µm[24] (vacuum wavelengths), which can be thermally populated for reasonable temperatures. Because the thermal equilibrium of the molecular gas can be driven by the energy distribution of the surface modes, it may happen that for simple polyatomic molecules, characterized by various vibration modes (*e.g.* $\nu_1$, $\nu_2$, $\nu_3$), with specific ladder of energy, only a single vibration mode (*e.g.* $\nu_2$) is efficiently populated, through its resonance in energy with the surface mode. This gas equilibrium, driven by the thermal emission in the surface near-field, can lead to a distribution of population, for the vibration –or rovibration– structure, in apparent violation of the Boltzmann energy distribution. This situation would just be a kind of "optical pumping by the surface modes". An experimental signature of such an unusual population distribution in the molecular levels may be searched for through standard gas spectroscopy (*e.g.* absorption). The only technical requirement is that transparent windows (at the frequency of the vibrations of interest), have to be inserted somewhere in the body of the cell. Indeed, the material whose near-field thermal emission governs the gas equilibrium, is intrinsically opaque in this frequency range.

## 5. CONCLUSION

The M-B distribution was successful at the onset of atomic theory, and was essential for the development of statistical physics. With precise velocity-selective spectroscopy, many hypotheses may be tested, notably at close distances from the surface. Although our dedicated experiment has not evidenced deviations to the M-B predictions, deviations should exist, for slow enough velocities. In some sense, the main surprise is that these deviations have not been looked for previously. This confidence in the M-B law has led to envision ultra-high resolution and "metrology-like" applications, through spectroscopy close to an interface, but fundamental obstacles may prevent such a realization. An unknown point is the ability of a dedicated set-up such as ours, to evidence deviations. The effective velocity distribution should also be assessed to allow a truly quantitative analysis of vapor spectra acquired at an interface, as the number –or distribution– of "slow" atoms can significantly affect the signal amplitude and lineshape in reflection spectroscopy, and thin cell spectroscopy as well.

At last, to emphasize the general interest of revisiting old problems, it is worth recalling that at Maxwell times, the nature of gas particles —and their electromagnetic interaction with the surface— was unknown : with this respect, an ideal gas was rather resembling a neutron gas[25]. Also, the problem of gas permeation and transport, often described through a Knudsen approach, remains very contemporary when dealing with atom size nanopores, like in graphene[26].

## ACKNOWLEDGMENTS

The work on the velocity distribution was supported by Bulgarian NSF contract DMU 02/17. The thin 60 µm Cs cell, and the high temperature all sapphire Cs cell were prepared by D. Sarkisyan group (Ashtarak, Armenia). The stay in France of J.C. de Aquino Carvalho for preparation of his PhD, was fully supported by CAPES (Brazil)